\documentclass[pre,preprint,eqsecnum]{revtex4}
\usepackage{bm}
\usepackage{graphicx}
\usepackage{amsmath}

\newcommand{\dosig}{\Omega_{ig}}
\newcommand{\dosex}{\Omega_{ex}}
\newcommand{\dosc}{\Omega_{config}}
\newcommand{\drmax}{\delta_{max}}
\newcommand{\dvmax}{\Delta_{max}}
\newcommand{\acc}{\frac{acc(A \to B)}{acc(B \to A)}}
\newcommand{\oprime}{\Omega'(E)}

\begin{document}

\title{On the Wang-Landau Method for Off-Lattice Simulations in the ``Uniform'' Ensemble}

\author{M. S. Shell}
\author{P. G. Debenedetti}
\email[Corresponding author; email: ]{pdebene@princeton.edu}
\author{A. Z. Panagiotopoulos}
\affiliation{Department of Chemical Engineering\\Princeton University, Princeton, NJ}
\date{\today}

\begin{abstract}
We present a rigorous derivation for off-lattice implementations of the so-called ``random-walk'' algorithm recently introduced by Wang and Landau [PRL \textbf{86}, 2050 (2001)].  Originally developed for discrete systems, the algorithm samples configurations according to their inverse density of states using Monte-Carlo moves; the estimate for the density of states is refined at each simulation step and is ultimately used to calculate thermodynamic properties.  We present an implementation for atomic systems based on a rigorous separation of kinetic and configurational contributions to the density of states.  By constructing a ``uniform'' ensemble for configurational degrees of freedom\textemdash in which all potential energies, volumes, and numbers of particles are equally probable\textemdash we establish a framework for the correct implementation of simulation acceptance criteria and calculation of thermodynamic averages in the continuum case.  To demonstrate the generality of our approach, we perform sample calculations for the Lennard-Jones fluid using two implementation variants and in both cases find good agreement with established literature values for the vapor-liquid coexistence locus.
\end{abstract}


\maketitle
\section{Introduction}

Computer simulations have become an important and well-established method for evaluating structural, dynamic, and equilibrium properties of substances.  In particular, Monte-Carlo (MC) methods in the canonical (constant $N, V, T$), isothermal-isobaric (constant $N, P, T$), and grand-canonical ensembles (constant $\mu, V, T$) are commonly used to obtain thermodynamic properties for given microscopic interactions \cite{1632, 1598}.  Though the conceptual basis for conventional MC simulations is straightforward, sampling constraints can emerge under certain circumstances.  Examples of such conditions include low-temperature and high-density systems for which ergodic sampling is difficult to achieve in a reasonable number of simulation steps.  As a consequence, numerous modifications of conventional MC methods have been proposed to enhance exploration of the phase space of a system.  Among these are annealing, parallel-tempering, and multicanonical algorithms (for an overview of these methods, see \cite{1598}.)

Recently, Wang and Landau proposed an elegant method for direct calculation of density of states in computer simulations \cite{1630, 931}.  The density of states, that is to say the degeneracy of energy levels available to the system, is directly related to entropy and can be used to calculate all thermodynamic properties at any conditions of interest \footnote{We should note that a single density of states function does not necessarily provide information about \emph{both} stable and metastable states.}.  In the Wang-Landau (WL) method, the probability of observing a particular atomic configuration is inversely proportional to the density of states corresponding to the given energy.  This sampling scheme ultimately results in a uniform distribution of macroscopic observables.  Though the density of states is not known \emph{a priori}, it is successively approximated by modification at each simulation step so as to ensure the uniform distribution.  The method is advantageous for two reasons.  First, a single, long simulation can provide information to calculate properties over a range of state conditions.  Second, the method appears to be affected less by the sampling problems of conventional MC simulations because energies are sampled with equal probability; this contrasts with conventional MC simulations for which high energy barriers are infrequently crossed.

As noted in \cite{1631}, the Wang-Landau method is most similar to the multicanonical techniques introduced by Berg and Neuhaus \cite{1601, 1615}.  (For a more thorough review of multicanonical methods and their application to fluid phase transitions, see \cite{1618}.)  Briefly, in a multicanonical simulation, one introduces an artificial sampling scheme that enhances the sampling of important states which are otherwise infrequently visited when the typical Boltzmann criterion is used.  This is particularly useful, for example, during a sub-critical grand canonical simulation when traversing the liquid-gas transition.  The sampling rule is constructed so that all macroscopic states are equally probable, i.e., it samples according to an inverse density of states.  Initially the density of states function is unknown.  It is generated iteratively over the course of several runs by maintaining histograms of states visited and updating between runs\textemdash frequently visited states are given higher values of the density of states.  At the end of the iterative procedure, a longer ``production'' simulation is performed.  True thermodynamic averages can then be generated by first unweighting the production results using the calculated density of states and then reweighting them with the Boltzmann rule. 

The Wang-Landau method also samples macroscopic states with equal probability.  Its distinguishing feature is the dynamic update of its acceptance rule; that is, the density of states estimate is modified at every simulation step rather than between runs.  This violates microscopic detailed balance; state probabilities fluctuate during the simulation.  The resolution of this violation is the following: over the course of the (long) simulation, the magnitude of the density of states modification is decreased until changes are just within the precision of the computer.  At this point, the detailed balance is \emph{essentially} satisfied.  Contrary to the microcanonical approach, the calculated density of states is not used to unweigh simulation results.  Rather, it is used directly via its statistical mechanical connection to entropy as given by Boltzmann's equation:
\begin{equation}
S=k \ln{\Omega(N,V,E)} \label{boltzmanns}
\end{equation}
where $k$ is Boltzmann's constant and $\Omega$ is the density of states.

In their original papers, Wang and Landau effectively applied their method to discrete systems.  For such cases, the complete set of energy levels can be enumerated and the density of states is stored, in exact form, as an array in the computer.  The use of the WL method in off-lattice systems, however, is emerging as an important simulation tool.  It has already been used successfully to describe properties of the Lennard-Jones fluid \cite{1631}.  Continuum systems require several nontrivial extensions of the original method.  For such systems, one must approximate the true density of states by a discretized version and choose, via trial-and-error or calculation, a finite range of energy over which to determine the density of states.  Furthermore, kinetic degrees of freedom, which are not explored during the simulation, must be taken into account in the processing of results.  

The rigorous connection between the WL approach in off-lattice systems and the actual density of states has not yet been addressed.  Specifically, the statistical-mechanical basis for developing acceptance criteria and for the treatment of kinetic degrees of freedom have not been systematically discussed in the literature to date.  Here we clarify the theoretical basis for the Wang-Landau method for continuum systems and discuss the logistics of its implementation.  We show that kinetic and configurational contributions to the density of states can be formally separated into the ideal gas and ``excess'' density of states functions, $\dosig$ and $\dosex$.  By casting the simulation in a ``uniform'' ensemble (i.e., one in which all macroscopic observables are equally probable), we derive the appropriate acceptance criteria and data analysis methods for simulations that probe $\dosex$.  Finally, we show that either of two types of simulation moves, involving particle number or volume fluctuations, may be used to explore the density dependence of $\dosex$ for single-component systems.  We believe the WL approach to be a powerful simulation algorithm, and so the aim of our derivation is to provide a starting point for future applications and extensions of the off-lattice version of the method.  

This paper is structured as follows. In section \ref{derivation}, we generalize the continuum WL method to the uniform ensemble and present the appropriate acceptance criteria and averaging procedures for simulations.  In section \ref{implementation}, we discuss several important numerical issues that arise in simulation of continuum systems, and in section \ref{casestudy} we compare results for the Lennard-Jones fluid using two variants of the method.

\section{Derivation of the method}
\label{derivation}

We begin with the classical microcanonical partition function for a single-component system of structureless particles, which may be written as
\begin{equation}
\Omega(N,V,E)=\frac{\epsilon}{h^{3N}N!}\int{\delta\left[E-H(p^{3N},q^{3N})\right]dp^{3N}dq^{3N}} \label{dos}
\end{equation}
where $N$ is the number of particles, $V$ is the volume, $E$ is the total energy, $h$ is Planck's constant, $\delta$ is the Dirac delta function, $p$ and $q$ are congugate momenta and positions, and $H$ is the Hamiltonian of the system.  The physical interpretation of the density of states is that $\Omega$ gives the number of states with energy $E$ accessible to $N$ particles in volume $V$.   The factor $\epsilon$ is a constant with units energy that characterizes the small energy interval into which the complete energy range is divided; its precise value, however, does not affect the calculation of thermodynamic quantities.  The factorial term and Planck's constant are quantum-mechanical in nature, the former accounting for the indistinguishibility of particles and the latter for the limit imposed by the uncertainty principle in the definition of a volume element in phase space.  It should be noted that in order to make \ref{dos} well-defined, we may replace $\delta$ with $\delta^{\Delta}$ which is a ``delta'' function of small but finite width $\Delta$ \cite{1634}.

Because $\Omega$ is known explicitly for an ideal gas, it is desirable to factor out the ideal gas density of states.  Since the Hamiltonian is a function of the kinetic and potential energies, the total system can be envisioned as separate kinetic and configurational subsystems that exchange energy.  Accordingly, equation \ref{dos} becomes
\begin{eqnarray}
\Omega(N,V,E) & = & \frac{\epsilon}{h^{3N}N!}
\int \delta\left[E-K(p^{3N})-U(q^{3N})\right]dp^{3N}dq^{3N}
\nonumber
\\
& = & \frac{\epsilon}{h^{3N}N!} 
\int 
\left\{ \int{\delta\left[E-t-K(p^{3N})\right]dp^{3N}} \right\}
\times \dots \nonumber \\
& & \dots
\left\{ \int{\delta\left[t-U(q^{3N})\right]dq^{3N}} \right\}
dt
\label{dossplit}
\end{eqnarray}
where $K$ is the kinetic energy, $U$ is the potential energy, $C$ is a normalization constant, and the outer integral is between the minimum energy the system can adopt and $E$.  In equation \ref{dossplit}, $\Omega$ is obtained by integrating over all possible distributions of energy between the kinetic and configurational subsystems.  This equation is rearranged to obtain the desired separation of ideal gas and configurational components:
\begin{eqnarray}
\Omega(N,V,E) & = & \epsilon^{-1} \int
\left\{ \frac{\epsilon V^{N}}{h^{3N}N!} \int{\delta\left[E-t-K(p^{3N})\right]dp^{3N}} \right\}
\times \dots \nonumber \\
& & \dots
\left\{ \frac{\epsilon}{V^{N}} \int{\delta\left[t-U(q^{3N})\right]dq^{3N}} \right\}
dt  
\nonumber
\\
& = & \epsilon^{-1} \int {
\dosig(N,V,E-t) \dosex(N,V,t)
dt }
\label{dossplit2}
\end{eqnarray}
where $\dosig$ is the ideal gas density of states, and $\dosex$ is the excess contribution due to configurational degrees of freedom.  Note that $\dosex$ is not the configurational density of states, which is given instead by
\begin{equation}
\dosc(N,V,E) = \frac{\epsilon}{q_0^{3N} N!} \int{ \delta \left[ E - U(q^{3N}) \right] dq^{3N} }
\label{dosconfig}
\end{equation}
where $q_0$ is a constant with units of length.  The relationship between $\dosc$ and $\dosex$ is
\begin{equation}
\dosex(N,V,E) \sim \frac{N!}{V^N}\dosc(N,V,E)
\label{dosnitodosc}
\end{equation}
where the trivial dependence on $q_0$ has been omitted.  In the Landau-Wang simulation methodology, sampling is performed according to the inverse density of configurational states, $\dosc$.  However, one may \emph{calculate} either $\dosex$ or $\dosc$ as long as the acceptance criteria and reweighting scheme are appropriately constructed (see section \ref{implementation}.)  In our simulations we choose to tabulate $\dosex$ rather than $\dosc$, mainly because excess properties have an intuitive physical basis.

Once one has calculated the excess contribution to the density of states, thermodynamic properties of interest are found by adding the ideal gas contribution.  The ideal gas density of states is given by
\begin{eqnarray}
\dosig(N,V,E) & = & \frac{\epsilon V^{N}}{h^{3N}N!} \int{\delta\left[E-K(p^{3N})\right]dp^{3N}}
\nonumber
\\
& = & { \left[ {\left( \frac{4\pi mE}{3h^2} \right)}^{3/2} \frac{Ve^{5/2}}{N^{5/2}} \right] }^N
\label{igdos}
\end{eqnarray}
where $m$ is the mass of the particles and $e$ is the base of the natural logarithm \cite{1589}.  (Several approximations have been made here, including the use of $\frac{3}{2}N-1 \approx \frac{3}{2}N$ and the assumption that $\epsilon$ is of negligible order.  For a detailed derivation, the reader is referred to \cite{1636}.)

The WL method can be generalized to a uniform ensemble for configurational degrees of freedom.  In this ensemble, the probability of observing a specific configuration is
\begin{equation}
P(N,V,q^{3N}) = \frac{1}{C} \frac{dq^{3N}dV}{\dosc \left( N,V,U(q^{3N}) \right)} 
\label{idosprob}
\end{equation}
where $q^{3N}$ represents the positions of the particles.  The normalization constant $C$ is given by 
\begin{equation}
C = \sum_{N=N_{min}}^{N_{max}} 
\int_{V_{min}}^{V_{max}} 
\left\{ \int_{E_{min}(N,V)<U<E_{max}(N,V)} \frac{1}{\dosc \left( N,V,U(q^{3N}) \right)} dq^{3N} \right\}
dV
\label{partitionfn}
\end{equation}
where the system potential energy, volume, and number of particles each varies between set bounds and the innermost multidimensional integral is over the system volume.  The existence of these limits implies that state probabilities are uniform within and zero outside of the specified range of $N, V, U$.

In the uniform ensemble, configurations characterized by specific values of $N, V, U$ will number according to $\dosc$, but will each have a probability inversely proportional to $\dosc$, resulting in a uniform distribution of energies, volumes, and numbers of particles (within the confines of the variable bounds.)  This is an extremely important property; it provides a feedback mechanism for calculating the density of states.  Based on deviations from a uniform distribution, we can systematically adjust an initial estimate for $\dosc$ (or $\dosex$) until we have converged on the true function to within the error of our adjustments.  The task is to design a simulation sampling scheme according to equation \ref{idosprob}.

In conventional single-component Monte Carlo simulations performed on spherically-symmetric particles, three types of moves are common: single particle displacements, volume scaling moves, and particle additions and deletions.  The acceptance criteria for these moves are derived by imposing a microscopic detailed balance that ensures equality of probability fluxes between pairs of states \cite{1598}.  For two states A and B, the acceptance criterion is formulated to yield
\begin{equation}
\frac{acc(A \to B)}{acc(B \to A)} = \left\{ P(B)\cdot\alpha(B \to A) \right\} \left\{ P(A)\cdot\alpha(A \to B) \right\}^{-1} .
\label{accprobformula}
\end{equation}
where $acc$ is the acceptance probability, $P$ is the equilibrium probability, and $\alpha$ is the Markov-chain transition probability.

For single particle moves, one selects a particle and makes a random displacement by an amount $-\drmax$ to $+\drmax$ in each component of its position.  Using equation \ref{accprobformula}, the detailed balance for this type of move is
\begin{eqnarray}
\acc & = & \left\{ \frac{(dq^{3N}dV)_B}{\dosc(N,V,U_B)} \cdot \frac{1}{N} \frac{(dq^{3})_A}{(2\drmax)^3} \right\}
\left\{ \frac{(dq^{3N}dV)_A}{\dosc(N,V,U_A)} \cdot \frac{1}{N} \frac{(dq^{3})_B}{(2\drmax)^3} \right\}^{-1}
\nonumber
\\
& = & \frac{\dosc(N,V,U_A)}{\dosc(N,V,U_B)}
\nonumber
\\
& = & \frac{\dosex(N,V,U_A)}{\dosex(N,V,U_B)}
\label{partmoveacc}
\end{eqnarray}
where the simplification in the second line arises from the fact that the differential elements $dV$ and the phase-space volume elements $dq^{3N}$ are equivalent in states A and B.   The third line results from the fact that the number of particles and volume remains constant.  In this move, the transition probabilities are symmetric, and thus cancel each other for constant $\drmax$.  It is not uncommon, however, to encounter a varying $\drmax$ in conventional Monte Carlo simulations, whereby $\drmax$ is dynamically changed to achieve a specified acceptance rate.  In this latter case, the detailed balance is not rigorously satisfied.  However, the distribution of sampled states in a conventional simulation is sharply peaked such that fluctuations away from the mean are small.  The result is that fluctuations in $\drmax$ are also small, and $\drmax$ is \emph{effectively} constant.  In the uniform ensemble, however, all states have equal probability and fluctuations from average values are large.  It is imperative, therefore, to explicitly maintain constant $\drmax$ after finding a good initial value.

For volume scaling moves, one increments the volume by an amount $-\dvmax$ to $+\dvmax$ and scales the entire simulation box and particle positions accordingly.  Contrary to the previous move, the phase-space volume elements $dq^{3N}$ are not equivalent in states of different volume.  The correct approach uses reduced coordinates $ds^{3N}=V^{-N}dq^{3N}$ which are equivalent across volumes. The acceptance criterion is
\begin{eqnarray}
\acc & = & \left\{ \frac{(V^Nds^{3N}dV)_B}{\dosc(N,V_B,U_B)} \cdot \frac{(dV)_A}{2\dvmax} \right\}
\left\{ \frac{(V^Nds^{3N}dV)_A}{\dosc(N,V_A,U_A)} \cdot \frac{(dV)_B}{2\dvmax} \right\}^{-1}
\nonumber
\\
& = & \frac{\dosc(N,V_A,U_A)}{\dosc(N,V_B,U_B)} \cdot \frac{V_B^N}{V_A^N}
\nonumber
\\
& = & \frac{\dosex(N,V_A,U_A)}{\dosex(N,V_B,U_B)}
\label{volmoveacc}
\end{eqnarray}
where, in the second line, the differential elements $dV$ and the reduced phase-space volume elements $ds^{3N}$ are equivalent in both states and cancel.  In this case, the transformation from an acceptance criterion involving $\dosc$ to $\dosex$ offers simplification.  Often, one would like to calculate thermodynamic properties over several orders of magnitude in volume, e.g., when investigating liquid-gas phase transitions.  Then, it becomes much more efficient to make volume scaling moves in the logarithm of the volume.  The acceptance criteria for this type of move is
\begin{eqnarray}
\acc & = & \left\{ \frac{(V^{N+1}ds^{3N}d\ln V)_B}{\dosc(N,\ln{V_B},U_B)} \cdot \frac{(d\ln V)_A}{2\ln\dvmax} \right\}
\left\{ \frac{(V^{N+1}ds^{3N}d\ln V)_A}{\dosc(N,\ln{V_A},U_A)} \cdot \frac{(d\ln V)_B}{2\ln\dvmax} \right\}^{-1}
\nonumber
\\
& = & \frac{\dosc(N,\ln{V_A},U_A)}{\dosc(N,\ln{V_B},U_B)} \cdot \frac{V_B^{N+1}}{V_A^{N+1}}
\nonumber
\\
& = & \frac{\dosex(N,\ln{V_A},U_A)}{\dosex(N,\ln{V_B},U_B)} \cdot \frac{V_B}{V_A}
\label{logvolmoveacc}
\end{eqnarray}
where we have switched to calculating $\Omega$ as a function of the logarithm of volume rather than the volume itself (this does not affect the behavior of $\Omega$ for a given volume.).  It should be noted that in both types of volume moves, the maximum volume change $\dvmax$ should remain constant during the production phase of the simulation, for the same reason mentioned for $\drmax$.

In particle addition or deletion moves, one inserts a particle at a random location or deletes a randomly chosen particle, respectively.  For the particle addition case, the detailed balance yields
\begin{eqnarray}
\acc & = & \left\{ \frac{(dq^{3N+3}dV)_B}{\dosc(N+1,V,U_B)} \cdot \frac{1}{N+1} \right\}
\left\{ \frac{(dq^{3N}dV)_A}{\dosc(N,V,U_A)} \cdot \frac{(dq^3)_B}{V} \right\}^{-1}
\nonumber
\\
& = & \frac{\dosc(N,V,U_A)}{\dosc(N+1,V,U_B)} \cdot \frac{V}{N+1}
\nonumber
\\
& = & \frac{\dosex(N,V,U_A)}{\dosex(N+1,V,U_B)}
\label{addpartacc}
\end{eqnarray}
where state B has one more particle than state A.  For a particle deletion,
\begin{eqnarray}
\acc & = & \left\{ \frac{(dq^{3N-3}dV)_B}{\dosc(N-1,V,U_B)} \cdot \frac{(dq^3)_A}{V} \right\}
\left\{ \frac{(dq^{3N}dV)_A}{\dosc(N,V,U_A)} \cdot \frac{1}{N} \right\}^{-1}
\nonumber
\\
& = & \frac{\dosc(N,V,U_A)}{\dosc(N-1,V,U_B)} \cdot \frac{N}{V}
\nonumber
\\
& = & \frac{\dosex(N,V,U_A)}{\dosex(N-1,V,U_B)}
\label{delpartacc}
\end{eqnarray}
where state B has one less particle than state A.  For both moves, the change from $\dosc$ to $\dosex$ results in criteria identical to those in equations \ref{partmoveacc} and \ref{volmoveacc}.

For the acceptance criteria just described, sampling can be performed using the traditional Metropolis algorithm:
\begin{equation}
acc(A \to B) = min \left( 1, \frac{\dosex(A)}{\dosex(B)} \right)
\label{metropolis}
\end{equation}
with the exception of equation \ref{logvolmoveacc}, for which there appear extra volume terms.  (The labels A and B have been used to abbreviate the values $N, V, U$ which characterize each configuration.)  Moves for which B is out of the range of the ensemble are rejected.  Initially, the density of states is given the value 1 everywhere; then, after each move during the simulation, its value at the current state is scaled.  If state C is the ending configuration after a move, being either A or B, the modification reads 
\begin{equation}
\left[ \dosex(C) \right]_{new} = f \cdot \left[ \dosex(C) \right]_{old}
\label{modifydos}
\end{equation}
where $f$ is a number greater than one, termed the modification factor.  The dynamic modification of the density of states in this way, coupled with the uniform ensemble, drives $\dosex$ to its true value to within a multiplicative constant.  It is important to recognize that the modification factor mediates the resolution of the calculated density of states.  If $f$ is large, the detailed balance is not satisfied and $\dosex$ will have large error fluctuations; when $f$ is very small, it will take an inordinate amount of simulation time to calculate $\dosex$.  The solution is to devise a schedule for the modification factor.  Initially $f$ is large, but in discrete steps at periods during the simulation, it is decreased until it approaches one.  (The details of this procedure are described in the following implementation section.)

For simulation purposes, it is important to recognize several properties of the density of states in the thermodynamic limit.  First, one only needs the intensive entropy for calculating thermodynamic properties, that is, $S/N = f(E/N,V/N)$.  Therefore, a simulation should make changes in energy density and particle density, of which the latter can be accomplished either by volume scaling moves \emph{or} particle additions and deletions.  Second, nearly all calculations of interest rely on derivatives of the entropy; therefore, entropy can be calculated to within an additive constant, i.e., $\dosex$ is known to a multiplicative constant. 

Once $\dosex$ has been generated, any thermodynamic property of interest can be calculated.  In principle, both equation \ref{dossplit2} and Boltzmann's equation could be used to determine the total density of states and, subsequently, thermodynamic properties from its various derivatives.  In practice, it is more convenient to average in an ensemble natural to the fluctuating quantities in the simulation.  This approach is especially important for properties which are sensitive to system size since it preserves the effects of the simulation fluctuations.  If volume scaling moves and particle displacements are used in the calculation of $\dosex$, the isobaric-isothermal ensemble is natural for calculations.  Given a pressure and temperature, the probability of a state in this ensemble is
\begin{equation}
P(V,K,U) = \frac{1}{\Delta(N,P,T)} \dosig(V,K) \dosex(V,U) e^{-(K+U+PV)/kT}
\label{probvkunpt}
\end{equation}
where $\Delta$ is the isothermal-isobaric partition function (which effectively normalizes the probabilities), $k$ is Boltzmann's constant and the dependence on $N$ in $\Omega$ has been suppressed (the number of particles is fixed in both our simulation and the ensemble.)  To determine the mean configurational energy $U$ and volume $V$ in this ensemble, one integrates \ref{probvkunpt} over volume, potential energy, and kinetic energy, the latter of which is analytic using equation \ref{igdos} for the ideal gas density of states:
\begin{eqnarray}
\overline{a}(P,T) & = & \int \int \int a(V,U) \cdot P(V,K,U) dUdKdV
\nonumber \\
& = & \int \int a(V,U) \cdot \left[ C(N,P,T) \dosex(V,U) e^{-(U+PV)/kT + N\ln{(V/V_0)}} \right] dUdV
\nonumber \\
& = & \int \int a(V,U) \cdot P(V,U) dUdV
\label{probvunpt}
\end{eqnarray}
where $a$ is either volume or potential energy. In the second line, we have substituted the ideal gas density of states and integrated over kinetic energy.  The constant $C(N,P,T)$ contains the result of this integration as well as the inverse partition function.  In practice, $C$ is calculated as the constant needed to normalize the probabilities given by the exponential.  Note that $V_0$ is an arbitrary reference volume to preserve units.  Its presence is aesthetic as its effects are eliminated by the normalization. 

A similar construction is made if the original simulation entails fluctuations in particle number and energy.  Here, the appropriate ensemble is the grand-canonical ensemble; the probability of a state given a temperature and chemical potential is
\begin{equation}
P(N,K,U) = \frac{1}{\Xi(\mu,V,T)} \dosig(N,K) \dosex(N,U) e^{-(K+U-\mu N)/kT}
\label{probnkumvt}
\end{equation}
where $\Xi$ is the grand-canonical partition function and the volume dependence of $\Omega$ has been suppressed.  The average potential energy and particle number are given by
\begin{eqnarray}
\overline{a}(\mu,T) & = & \sum_N \int \int a(N,U) \cdot P(N,K,U) dUdK
\nonumber \\
& = & \sum_N \int a(N,U) \cdot \Bigg[ \Xi^{-1} \dosex(N,U) e^{-(U-\mu N)/kT-\frac{5}{2} N\ln{N}+\mu_0(V)N/kT} 
\times \dots \nonumber \\
& & \dots \left( \int K^{\frac{3}{2}N} e^{-K/kT} dK \right) \Bigg] dU
\nonumber \\
& = & \sum_N \int a(N,U) \cdot \Bigg[ \Xi^{-1} \dosex(N,U) e^{-(U-\mu 'N)/kT-\frac{5}{2}N\ln{N}} 
\times \dots \nonumber \\
& & \dots \left( \frac{kT}{e(\frac{3}{2}N+1)} \right)^{\frac{3}{2}N+1} \Bigg] dU
\nonumber \\
& = & \sum_N \int a(N,U) \cdot \left[ \Xi^{-1} \dosex(N,U) e^{-(U-\mu 'N)/kT-N\ln{N}+\frac{3}{2}N\ln{\frac{kT}{E_0}}} \right] dU
\nonumber \\
& = & \sum_N \int a(N,U) \cdot P(N,U) dU
\label{probnumvt}
\end{eqnarray}
where $a$ is either the number of particles or potential energy.  In the second line, we substitute the ideal gas density of states and let the volume-dependent term $\mu_0$ contain all terms in the exponential which are linear in $N$.  This simply serves to shift the zero of the chemical potential, reflected in subsequent lines with the notation $\mu'$.  In the third and fourth lines, we integrate over the kinetic energy, in which we use Stirling's formula and make the approximation $\frac{3}{2}N+1 \approx \frac{3}{2}N$.  The constant $E_0$ is again introduced to conserve units; its effect is absorbed in $\mu'$.  As in the previous case, $\Xi$ is calculated in the process of normalizing the probability $P(N,U)$.

Of particular interest in simulation work is the prediction of phase transitions.  Under state conditions favoring a two-phase system, the joint probabilities $P(V,U)$ and $P(N,U)$ will appear bimodal; phase equilibrium occurs when the probability volume under the two peaks is equal.  In practice, often one identifies some intermediate $V_{mid}$ or $N_{mid}$ which separates the two peaks, sets the field parameter constant (pressure or chemical potential), and adjusts the temperature until the probability volumes are equal.  This only works well at sub-critical conditions where the probability of observing the intermediate density is extremely low.  Near the critical point, finite-size scaling methods are more useful (not discussed here; see \cite{1219} for example.)  The condition of equality of probability volumes in the isothermal-isobaric and grand canonical cases becomes, respectively,
\begin{eqnarray}
\int_{V<V_{mid}} \int P(V,U)dUdV & = & \int_{V>V_{mid}} \int P(V,U)dUdV
\label{nptequilib}
\\
\sum_{N<N_{mid}} \int P(N,U)dU & = & \sum_{N \ge N_{mid}} \int P(N,U)dU
\label{mvtequilib}
\end{eqnarray}
where the dependence of $P$ on temperature and pressure or chemical potential is implicit.  Once conditions for phase equilibrium are determined, equations \ref{probvunpt} and \ref{probnumvt} can be used to determine properties of a specific phase by restricting the integrals to the phase's density range.  For example, in the isothermal-isobaric case,
\begin{eqnarray}
\overline{a}^{I} & = & \int_{V<V_{mid}} \int a(V,U) \cdot P(V,U) dUdV
\nonumber \\
\overline{a}^{II} & = & \int_{V>V_{mid}} \int a(V,U) \cdot P(V,U) dUdV
\label{phaseaverages}
\end{eqnarray}
where the superscript numerals indicate the phase.  

\section{Implementation of the method} \label{implementation}

Attention must be paid to several issues when implementing the WL method in a simulation.  The most obvious problem is the calculation of $\dosex$, which can span many orders of magnitude and quickly pose over/underflow precision problems for the computer.  Following Wang and Landau \cite{1630}, we tabulate $\ln\dosex$ rather than $\dosex$ itself and make modifications of the type $\ln\dosex = \ln\dosex + \ln{f}$ where $f$ is our modification factor.  Accordingly, the acceptance criterion in equation \ref{metropolis} becomes
\begin{equation}
acc(A \to B) = min \bigg( 1, \text{exp}[ \ln\dosex(A) - \ln\dosex(B)] \bigg) .
\label{metropolis2}
\end{equation}

The WL method was originally implemented for discrete systems \cite{1630, 931}.  In this case, the density of states is a discrete function which can be tabulated as an array in the computer.  For continuum systems, as noted in \cite{1631}, it is first necessary to discretize the density of states function in energy and, if volume scaling moves are being performed, volume as well.  The degree of discretization that is necessary to obtain accurate results is not straightforward; if the grid is treated as a linear approximation, for example, enough bins must be used to capture the curvature of the entropy surface being investigated.  

It is useful to perform energy and volume interpolation on the grid used for the density of states.  Without interpolation, a system may be able to stay within a specific grid level for large numbers of simulation steps.  With it, the level corresponding to such a ``stagnant'' series of configurations will effectively develop a sharp peak at its center as a result of the modification factor; this motivates the system leaving that level.  We use bilinear interpolation for our simulations.  We should note that interpolation should be used only when all neighboring grid points are well-defined, that is, when all are in an energy-accessible range. 

In order to determine the accessible energy range at each density, it is necessary to first carry out a small set of simulations.  We perform a short Monte Carlo NVT simulation for each discretized density at the lowest temperature we are interested in studying.  In doing so, we make note of the lower-bound potential energy sampled during these simulations and form a border in density space of energies below which we do not attempt to calculate the density of states (i.e., we reject moves outside of the border.)  Without this step, we find that the simulation can get trapped for large numbers of simulation steps in states of very low degeneracy.  

During the subsequent ``production'' phase of the simulation, the schedule of changes in the modification factor affects the quality of the calculated density of states.  In the original method, Wang and Landau use a histogram of states as a signal for these changes.  They start $f$ at a large value ($\ln{f}=1$), and run the simulation until a flat histogram is achieved, i.e., until they observe a uniform distribution of states.  Then they decrease the modification factor according to the rule $\ln{f_{new}}=\frac{1}{2}\ln{f_{old}}$ and repeat the procedure until $f$ is near one ($\ln{f}=10^{-8}$.)  As the authors discussed, this approach is only mildly satisfactory since there is still arbitrariness in developing a criteria for the ``flatness'' of the histogram.  We choose instead to require that each discrete state be visited a minimum number of times before changing the modification factor (e.g., 20 times.)  Though this may also seem arbitrary, it does guarantee that each value in the density of states will have a chance to adjust to the resolution of the current modification factor.

In the original Wang-Landau work, it was noted that one needs to perform several independent simulations for regions of large entropy gradients \cite{1630}.  That is, the total energy range is divided and the density of states is calculated for each section.  Then, the density of states for the whole range is obtained from those of the sections by matching values in overlapping energy regions (one adjusts the multiplicative constants to which each density of states is known.)  If such a procedure is not implemented, the number of simulation steps necessary for complete coverage of the total energy range can become extremely large.  This is analogous to long ``ergodicity times'' or ``tunneling times'' in multicanonical methods \cite{1613, 1615}.  

We use the same approach in our implementation by creating subsections of the master energy and density range studied and running a separate simulation for each.  Since the error in our calculated $\ln\dosex$ functions is essentially the same at every point and proportional to the modification factors we used, we shift each $\ln\dosex$ to minimize the total variance in regions of overlap between the subsections.  The total error is defined as
\begin{equation}
e_{tot} = \sum_{i=1}^{N} \sum_{j=i+1}^{N} \sum_{k} \left( \ln{\Omega_i(k)} + C_i - \ln{\Omega_j(k)} - C_j \right)^2
\label{errortotal}
\end{equation}
where $N$ is the number of subsections, $k$ is an index for all overlapping discretized points in $\ln\Omega$ of the two subsections $i$ and $j$, and the constants $C$ are the values by which we shift.  In this equation, we consider only overlap between pairs of subsections.  In minimizing $e_{tot}$ with respect to the constants $C$, we obtain $N$ equations.  One $C$ value must be specified to obtain a solution; we solve the remaining $N-1$ equations using a matrix inversion algorithm.  Once we know the shifting values, the final density of states is then pieced together from each shifted subsection; values at areas of overlap are averaged.

There is one caveat associated with using energy and density subsections in the uniform ensemble: each subsection must have a sufficient range of energy to allow room for sampling all relevant configurations.  That is, there must be an adequate number of paths for the system to move between densities; otherwise, it is very difficult to obtain good convergence of the calculated density of states.  This amounts to setting the maximum potential energy of each subsection to a sufficiently large value.  In the majority of our simulations, we find that creating subsections that differ only in density range (i.e., have the same range of potential energy) is the most effective approach.  Though the undivided energy range can contain significant entropy gradients, creating energy subsections can actually result in longer runs because the system's ergodicity is restricted.

\section{Case Study: The Lennard-Jones Fluid}
\label{casestudy}

It was shown in \cite{1631} that the phase-behavior of the Lennard-Jones fluid is well reproduced by the WL algorithm with particle displacement and addition/deletion moves.  Here we generalize those results to show that the formalism we have outlined is extensible to all types of simulation moves.  We conduct simulations of particles interacting through the cut and long-range corrected Lennard-Jones potential with a cutoff radius of 2.5$\sigma$.   Particle displacement moves are used with two cases of density changes: volume scaling moves in the logarithm of the volume and particle addition/deletion moves.  For the volume scaling case, we use a system of 128 particles and allow the box width to vary between 5.04$\sigma$ and 21.6$\sigma$, corresponding to reduced densities between 1.0 and 0.013.  The density of states function is discretized into 500 energy and 200 volume bins, with an energy range of -806$\epsilon$ to 64$\epsilon$.  For the particle addition/deletion case, we use a box of width 5$\sigma$ and allow the number of particles to fluctuate between 2 and 111, corresponding to densities between 0.016 and 0.89.  The energy range of the density of states is divided into 1000 bins and spans -700$\epsilon$ to 20$\epsilon$.  

In both cases, we start our simulations with $\ln{f}=1$ and require that each discretized point in the density of states be visited 20 times before the modification factor is updated.  The update is performed according to $\ln{f_{new}}=\frac{1}{2}\ln{f_{old}}$.  We stop the simulation when $\ln{f}<10^{-5}$.  Approximately 120 hours on an AMD Athlon 1.4 GHz workstation are required for the complete simulation; however, nearly 100 of those hours correspond to $\ln{f}<10^{-3}$ for which the density of states is already reasonably converged.  For comparison, the same potential code was used to generate two-dimensional histograms of energy and particle number in grand-canonical MC simulations.  Data from seven state points near coexistence was obtained by long production runs totaling 25 hours of computer time.  The resulting histograms are of high enough quality to determine the entropy in the region of phase coexistence using histogram reweighting techniques (see \cite{1219} for details of such a procedure.) 

\begin{figure}
\includegraphics{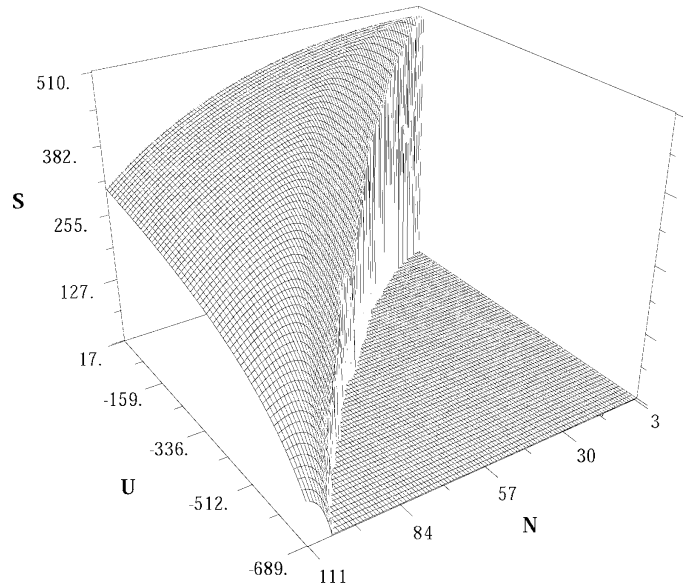}
\caption{\label{molgibbs}Gibbs surface for Lennard-Jones excess entropy as calculated from particle displacement and addition/deletion moves.}
\end{figure}
\begin{figure}
\includegraphics{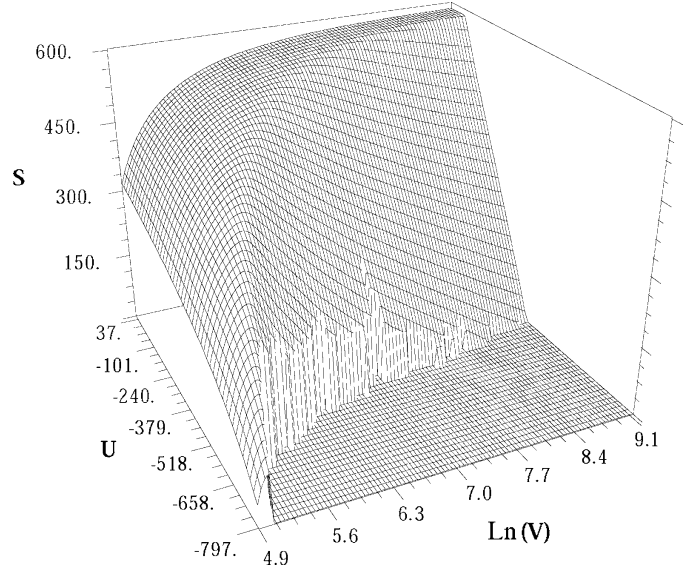}
\caption{\label{volgibbs}Gibbs surface for Lennard-Jones excess entropy as calculated from particle displacement and volume scaling moves.  The irregular low-energy boundary is the result of variations in calculating the border for lowest accessible energies at each density.}
\end{figure}

The Gibbs surfaces for the excess entropy calculated in the two cases are shown in Figures \ref{molgibbs} and \ref{volgibbs}.  It is apparent from Figure \ref{molgibbs} that the accessible range of energy in the particle addition/deletion case is extremely sensitive to the number of particles.  At very low particle numbers, the number of discretized points in the density of states which have an accessible energy is small.  Furthermore, at these small particle numbers there is a sharp peak in the excess entropy at an intermediate energy, requiring a greater number of interpolation bins to be reproduced accurately (an explanation of this peak is below).  This necessitates a high degree of discretization in the particle addition/deletion case, which can unfavorably increase the duration of the simulation.  In contrast, the accessible range of energy does not vary drastically with density in the volume scaling moves case.  The ability of particles in a larger volumes to condense into a droplet results in a low-energy ``tail'' which extends the energy range of low-density configurations.  Thus, we find the volume scaling approach to be advantageous in the calculation of excess entropy.

\begin{figure}
\includegraphics{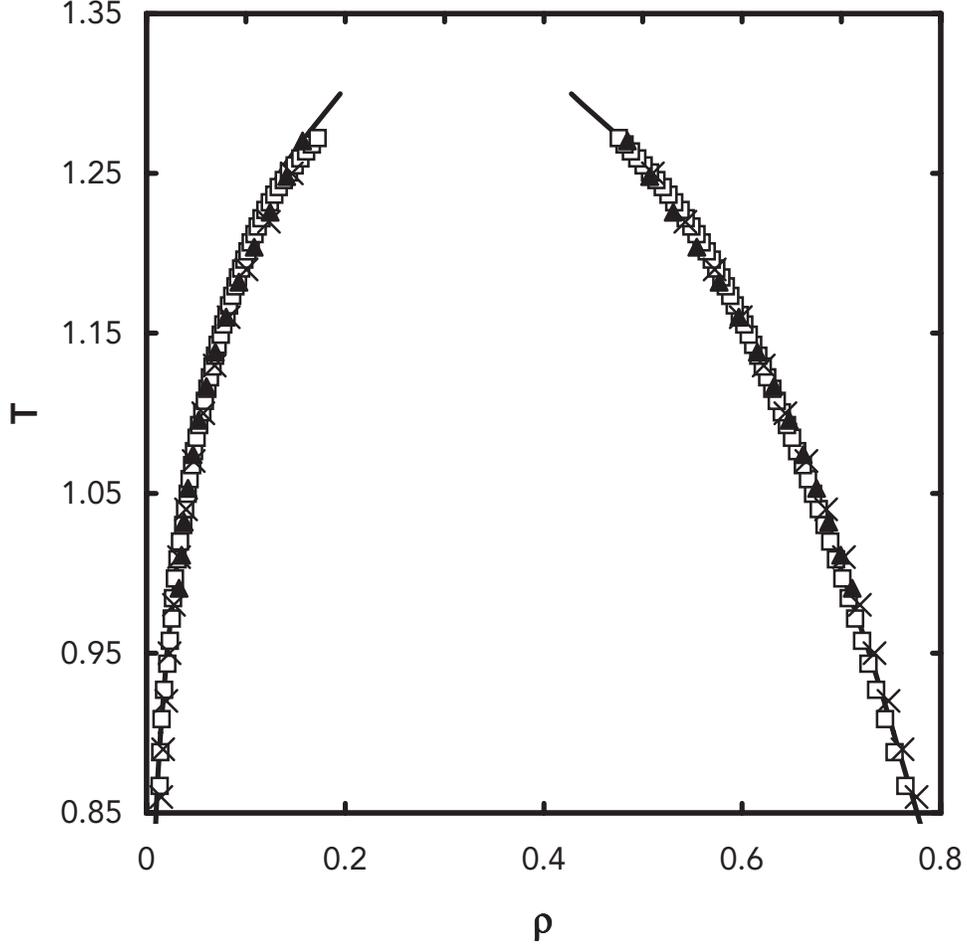}
\caption{\label{ljphase}Vapor-liquid equilibria for the Lennard-Jones fluid.  Squares and triangles correspond to results from the volume scaling and particle addition/deletion variants of the off-lattice Wang-Landau method, respectively.  Crosses are from a histogram reweighting study using grand-canonical MC data at seven state points.  The solid line is from Lotfi, et. al. \cite{1635}. }
\end{figure}

The vapor-liquid equilibrium data as calculated from the two simulation variants are shown in Figure \ref{ljphase} and compared to literature values.  Both results are truncated just below the critical point, where finite-size effects become significant.  Additionally, results from the particle addition/deletion case must be truncated around $T=1.0$ reduced temperature.  Below this temperature, the calculations are strongly influenced by fact that we have limited the minimum number of particles to two.  For the reported data in each case, we obtain good agreement with literature values.  

\begin{figure}
\includegraphics{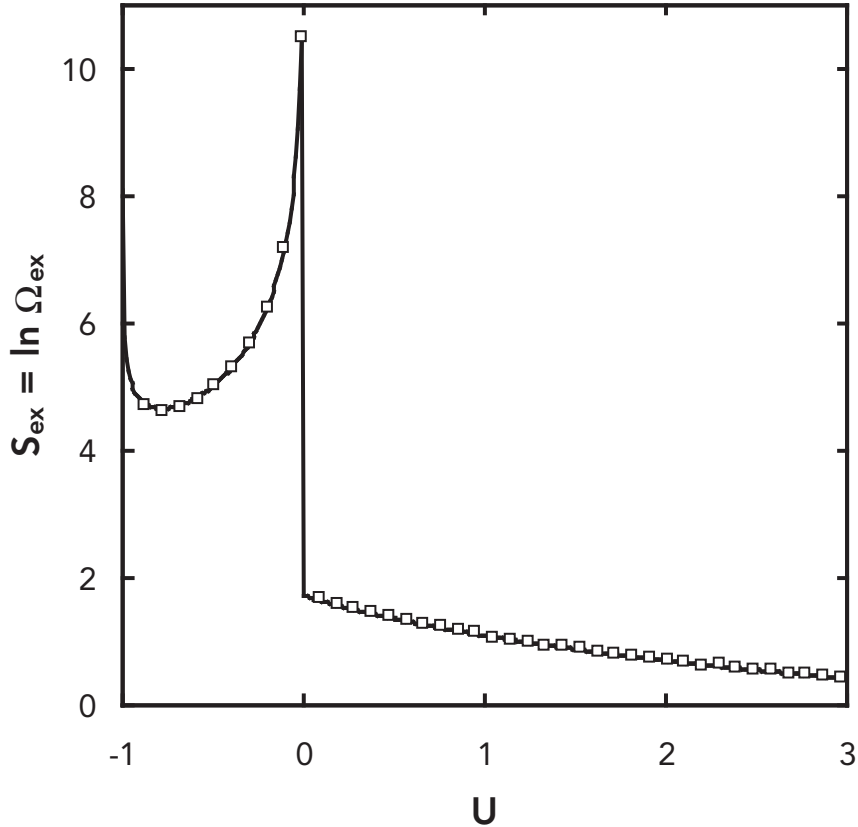}
\caption{\label{lj2atoms}Calculated and analytical excess entropy for two Lennard-Jones particles in a box of side length 5$\sigma$.  Points are simulation results and the line is from the analytical calculation.}
\end{figure}

As a further verification of the method, we also compare calculated and analytical values for the excess entropy of the two-particle Lennard-Jones system in a box of width 5$\sigma$.  The analytical values are found by placing one of the particles at the center of the simulation box, and finding $\oprime$, the total volume in which the second particle can be placed such that the energy is less than $E$.  The result is
\begin{eqnarray}
\oprime = \left\{ \begin{array}{ll}
\frac{4\pi}{3} \left\{ \left[ \frac{1}{2}-\frac{1}{2} \left( 1+E \right)^{\frac{1}{2}} \right]^{-\frac{1}{2}} - 
\left[ \frac{1}{2}+\frac{1}{2} \left( 1+E \right)^{\frac{1}{2}} \right]^{-\frac{1}{2}} \right\}
\ \ \ & E \le E^*  \\
8 H^3  - \frac{4\pi}{3} \left[ \frac{1}{2}+\frac{1}{2} \left( 1+E \right)^{\frac{1}{2}} \right]^{-\frac{1}{2}} 
\ \ \ & E > 0
\end{array} \right.
\label{2atom}
\end{eqnarray}
where $H$ is half of the box width and $E^*$ is the interaction energy when the particles are separated by $H$.  (For legibility, we have omitted in this presentation the region $E^* < E < 0$ which entails the calculation of the intersection volume for a sphere and cube.)  The excess entropy, given to within an additive constant, is the logarithm of the derivative of $\Omega'(E)$ with respect to E.  In Figure \ref{lj2atoms} we show this calculation alongside the results of the WL simulation algorithm.  The agreement is quite good.  We should note that the odd shape of this curve is the result of the small system size.  The peak around $E=0$ corresponds to the large amount of possible locations for the second particle which are relatively far from the first (i.e., which result in near-zero interaction energy.)  As one increases in energy for $E>0$, the excess entropy decreases; here the second particle must be placed in a volume which is essentially a shrinking shell around the first.  It is interesting to note that this region corresponds to a state of negative configurational temperature in the two-particle system.

\section{Conclusion}
\label{conclusion}

We have shown that the Wang-Landau algorithm for calculating the density of states can be generalized to continuum systems using the so-called ``uniform'' ensemble.  This ensemble permits calculation of the excess density of states, a function that measures the degeneracy of energy levels due to configurational degrees of freedom.  Thermodynamic properties are found from the connection between the density of states and entropy; in practice, they are determined by averaging in an ensemble appropriate to the type of Monte-Carlo moves used in a particular application.  For simulation purposes, we have derived acceptance criteria for particle displacement, volume scaling, and particle addition/deletion moves, though the uniform ensemble can be applied to any Monte-Carlo simulation move.  For single-component systems, either volume scaling or particle addition/deletion moves can be used to explore the density dependence of the excess density of states.  In the case of the Lennard-Jones fluid, we find the former has advantages at low density.  

Though the Wang-Landau algorithm is conceptually elegant, it does not offer a significant time-saving advantage over comparable methods for the calculation of liquid-gas equilibria (e.g., multicanonical or histogram reweighting techniques.)  Its primary benefit is that it makes no reference to temperature; its sampling scheme has the unphysical advantage that high energy barriers are sampled with the same probability as low-energy configurations.  This makes the WL method particularly attractive for low temperature studies, and has the potential to provide new and reliable data about the equilibrium behavior of supercooled liquids, glasses, and polymers for which simulation time scales have previously been prohibitive.  Studies have begun to demonstrate its potential usefulness in such applications, including protein folding \cite{1637} and polymer films \cite{1638}.

\begin{acknowledgements}
We gratefully acknowledge the support of the Department of Energy, Division of Chemical Sciences, Geosciences, and Biosciences, Office of Basic Energy Science (grants DE-FG02-87ER13714 to PGD and DE-FG02-01ER15121 to AZP.)  MSS is further thankful for the fellowship support of the Fannie and John Hertz Foundation. 
\end{acknowledgements}

\bibliography{idos-BibTeXExport}

\end{document}